\documentclass[epj]{svjour}
\usepackage{graphicx,epsfig}
\usepackage{amsmath,amssymb}
\usepackage{array}
\usepackage{color}
\usepackage{cite}

\newcommand{\be}{\begin{equation}}
\newcommand{\ee}{\end{equation}}
\newcommand{\ba}{\begin{array}}
\newcommand{\ea}{\end{array}}
\newcommand{\bea}{\begin{eqnarray}}
\newcommand{\eea}{\end{eqnarray}}

\newcommand{\gsim}{\stackrel{>}{_\sim}}

\newcommand{\cO}{{\cal O}}
\newcommand{\cB}{{\cal B}}

\newcommand{\BR}{{\cal B}}
\newcommand{\cL}{{\cal L}}

\newcommand{\no}{\nonumber}
\newcommand{\Btaun}{{B \to \tau \nu}}
\newcommand{\epsK}{\varepsilon_K}
\newcommand{\realV}{V}

\begin{document}

\title{$B \rightarrow \tau \nu$ in Two-Higgs doublet models with MFV}

\author{Gianluca Blankenburg\inst{1}, 
Gino Isidori\inst{2}
}

\institute{Dipartimento di Fisica, Universit\`a di Roma Tre,  Via della Vasca Navale 84, I-00146 Roma, Italy
\and
INFN, Laboratori Nazionali di Frascati, Via E.~Fermi 40, I-00044 Frascati, Italy}

\authorrunning{G.~Blankenburg and G.~Isidori }
\titlerunning{$B \rightarrow \tau \nu$ in Two-Higgs doublet models with MFV}
\mail{gino.isidori@lnf.infn.it}

\date{Received: date / Revised version: date}

\abstract{We analyse the $B \rightarrow \tau \nu$ decay in a generic two-Higgs doublet 
model satisfying the MFV hypothesis. 
In particular, we analyse under which conditions $\BR(B \rightarrow \tau \nu)$ can be
substantially enhanced over its SM value, taking into account the constraints of 
$K \rightarrow \mu\nu$, $B \rightarrow X_s \gamma$, and $B_s \rightarrow \mu^+ \mu^-$.
We find that for large $\tan\beta$ values and Peccei-Quinn symmetry breaking terms
of $\cO(1/\tan\beta)$ a sizable ($\sim 50\%$) enhancement of  $\BR(B \rightarrow \tau \nu)$
is possible, even for $m_H \sim 1$~TeV. 
\PACS{{13.20.He}{12.60.-i}}
}

\maketitle

\section{Introduction}

The leptonic decays $B\to \ell\nu$ are particularly interesting 
probes of the Higgs sector and, particularly, of the Yukawa interaction.
%
The $\tau$  channel is the only decay mode of this type observed so far.
The experimental world average~\cite{Barlow:2011fu},
\be
\cB(B\to \tau \nu)^{\rm exp} = (1.64\pm 0.34)\times 10^{-4}~,
\label{eq:exp}
\ee
has to be compared with the SM prediction
\be
\cB(B\to \tau \nu)^{\rm SM} = 
\frac{G_{F}^{2}m_{B}m_{\tau}^{2}}{8\pi}\left(1-\frac{m_{\tau}^{2}}
{m_{B}^{2}}\right)^{2}f_{B}^{2}|V_{ub}|^{2}\tau_{B}~,
\label{eq:SM}
\ee
whose uncertainty is mainly due to the determination of $|V_{ub}|$
and $f_B$. Using the best fit values of $|V_{ub}|$ from global CKM fits,
the  UTfit~\cite{Bevan:2010gi} and CKMfitter~\cite{CKMfitter} collaborations quote
\bea
\cB(B\to \tau \nu)^{\rm SM} &=& (0.79 \pm 0.07)\times 10^{-4}~[{\rm UTfit}]~, 
\nonumber \\
\cB(B\to \tau \nu)^{\rm SM} &=& (0.76~{}^{+~0.10}_{-~0.06})\times 10^{-4}~[{\rm CKMfit}]~.
\nonumber 
\eea
These low values correspond to a $2.5 (2.8)~\sigma$ deviation from 
the experimental result in Eq.~(\ref{eq:exp}).
Motivated by this discrepancy, we analyse in this paper 
the predictions of $B\to \tau \nu$ in models with 
an extended Higgs sector.

The helicity suppression in Eq.~(\ref{eq:SM})
is a manifestation of the key role played by the 
Yukawa interaction in the $B\to \tau \nu$ decay amplitude. 
For instance, in models with two Higgs
doublets coupled separately to up- and down-type quarks
(2HDM-II models), the $B\to \tau \nu$ amplitude receives an additional 
tree-level contribution from the heavy charged-Higgs exchange, 
leading to~\cite{Hou:1992sy}
\be
\frac{\BR^{\rm 2HDM-II}(\Btaun)}{\BR^{\rm SM}(\Btaun)} \no\\
=\left[1- \frac{m^{2}_B \tan^2\beta }{m^{2}_{H}} \right]^2~,
\label{eq:Btn0}
\ee
where $\tan\beta=v_2/v_1$
is the ratio of the two Higgs vacuum expectation values
and $m_{H}$ is the charged-Higgs boson mass.
For large $\tan\beta$ values the ratio in (\ref{eq:Btn0}) can be 
substantially different from one. However, within this simple
framework the interference sign of SM and non-standard 
amplitudes is fixed. Taking into account the constraints on $m_{H}$ 
from other processes, this sign implies a suppression 
of $\BR(\Btaun)$ in the 2HDM-II compared to the SM, worsening 
the comparison with the experimental result in Eq.~(\ref{eq:exp}).
The sign of SM and non-SM contributions can be varied 
in generic models with new sources of flavour-symmetry breaking.
However, in this case the problem is how to justify the good 
agreement between SM predictions and observations in most 
other flavour-violating observables. 

A consistent assumption usually invoked to avoid large deviations from the SM
in flavour-violating processes is the so-called Minimal Flavor Violation (MFV) 
hypothesis~\cite{D'Ambrosio:2002ex} (see also~\cite{Chivukula:1987py,Buras:2000dm,Kagan:2009bn}).
According to this hypothesis, the quark Yukawa couplings are the 
only sources of quark flavour-symmetry breaking 
both within and beyond the SM.
As recently discussed in~\cite{Buras:2010mh,Buras:2010zm},
the MFV hypothesis applied to two-Higgs doublet
models not only provides a sufficient protection of FCNCs 
in case of extra scalar degrees of freedom, it can also 
provide an explanation of the existing tensions 
in $\Delta F=2$ observables. More explicitly, it has been show that 
two-Higgs doublet models respecting the MFV hypothesis with the 
inclusion of flavour-blind CP-violating (CPV) phases
(dubbed $\text{2HDM}_{\overline{\text{MFV}}}$ framework), 
can accommodate a large CPV phase in $B_s$ mixing  
softening in a correlated manner the observed anomaly
in the relation between $\epsK$ and $S_{\psi K_S}$~\cite{Buras:2010mh}.
In this work we analyse under which conditions, 
within the  $\text{2HDM}_{\overline{\text{MFV}}}$ framework, 
$\cB(B\to \tau \nu)$ can be enhanced 
over it SM prediction.\footnote{~Recent analyses 
of $B\to \tau \nu$ in different models with 
an extended Higss sector 
can be found in Ref.~\cite{Dobrescu:2010rh,Jung:2010ik,Bhattacherjee:2010ju}.}

\section{The $\text{2HDM}_{\overline{\text{MFV}}}$ framework}

We consider a model with two Higgs fields, $H_{1,2}$, with opposite 
hypercharge ($Y=\pm 1/2$). The generic form of the Yukawa interaction 
for such a Higgs sector is
\bea
- \cL_Y^{\rm gen} &=& 
\bar Q_L X_{d1} D_R H_1 + \bar Q_L X_{u1} U_R H_1^c
\nonumber \\
&+& \bar Q_L X_{d2} D_R H_2^c + \bar Q_L X_{u2} U_R H_2 +{\rm h.c.}~,
\label{eq:generalcouplings}
\eea
where $H_{1(2)}^c=-i\tau_2 H_{1(2)}^*$.
The two real vacuum expectation values (vevs) are
defined as $\langle H^\dagger_{1(2)} H_{1(2)} \rangle=v^2_{1(2)}/2$, 
with $v^2 =v_1^2+v_2^2 \approx (246~{\rm GeV})^2$,
and, as anticipated, $\tan\beta=v_2/v_1$.

The $X_i$ are $3\times 3$ matrices in flavour space.
The general structure implied by the MFV hypothesis for these 
matrices is a polynomial expansions in terms of the two
(left-handed) spurions $Y_u Y_u^\dagger$ and $Y_d Y_d^\dagger$~\cite{D'Ambrosio:2002ex,Buras:2010mh}:
\bea
X_{d1} &\equiv& Y_d~, \nonumber \\
X_{d2} &=& \epsilon_{0} Y_d + \epsilon_{1} Y_d  Y_d^\dagger Y_d
        +  \epsilon_{2} Y_u Y_u^\dagger Y_d + \ldots~, \nonumber \\
X_{u2} &\equiv& Y_u~, \nonumber \\
X_{u1} &=& \epsilon^\prime_{0} Y_u + \epsilon^\prime_{1}  Y_d Y_d^\dagger Y_u
+  \epsilon^\prime_{2}  Y_u Y_u^\dagger Y_u + \ldots~,
\label{eq:XMFVgen}
\eea
where the $\epsilon^{(\prime)}_{i}$ are complex parameters. 
We work under the assumption $\epsilon^{(\prime)}_{i} \ll 1$,
as expected by an approximate $U(1)_{\rm PQ}$ symmetry
that forbids non-vanishing $X_{u1,d2}$ at the tree level, 
and we assume $\tan\beta =t_\beta = s_\beta/c_\beta \gg 1$.
For simplicity, we also restrict the attention to terms with 
at most three Yukawa couplings in this expansion (namely we consider 
only the terms explicitly shown above) and we assume real 
$\epsilon^{(\prime)}_{i}$ since we are interested only in CP-conserving
observables.  Finally, we assume negligible violations of the $U(1)_{\rm PQ}$
symmetry in the lepton Yukawa couplings.

After diagonalizing quark mass terms and rotating the Higgs fields such that only one
doublet has a non-vanishing vev, the interaction of down-type quarks with the neutral
Higgs fields assumes the form
\be
\cL_{\rm n.c.}^{d} = -  \frac{\sqrt{2}}{v} \bar d_L M_d d_R \phi_v^0
- \frac{1}{s_\beta}{\bar d}_L Z^{d}
\lambda_d d_R \phi_H^0 {\rm +h.c.}\,,
\label{eq:LH_FCNC}
\ee
where $\phi_v$ ($\phi_H$) is the linear combination of $H_{1,2}$
with non-vanishing (vanishing)
vev $\langle\phi^0_v\rangle =v/\sqrt{2}$ ($\langle\phi^0_H\rangle =0$).
The flavour structure of the $Z^d$ couplings is
$$
Z^d_{ij} =
\bar{a}\delta_{ij} +
\left[
a_0 V^\dagger \lambda_u^2 V + a_1 V^\dagger \lambda_u^2 V
\Delta + a_2 \Delta V^\dagger \lambda_u^2 V
\right]_{ij}\,,
$$
where $V$ is the physical CKM matrix, $\Delta \equiv \rm{diag}(0,0,1)$, $\lambda_{u,d}$
are the diagonal Yukawa couplings in the limit of unbroken 
$U(1)_{\rm PQ}$ symmetry, and the $a_i$ are flavour-blind coefficients 
(see~\cite{D'Ambrosio:2002ex,Buras:2010mh} for notations).
Similarly, the interaction of the quarks with the physical charged Higgs is
described by the following flavour changing effective Lagrangian~\cite{D'Ambrosio:2002ex}
\be
\cL_{\rm H^+ } = \left[
 {\bar U}_L C_{R}^{H^+} \lambda_d D_R  
+  \frac{1}{t^2_\beta} {\bar U}_R  \lambda_u  C_{L}^{H^+} D_L  \right] H^+  {\rm +h.c.},
\label{eq:LH_charged}
\ee
where the flavour structure of $C_{L,R}^{H^+}$ is
\bea
C_{R}^{H^+} &=& \left( b_0 \realV + b_1 \realV \Delta + b_2 \Delta\realV + b_3 \Delta\right)~,
\label{eq:C_R^H+} \\
C_{L}^{H^+} &=& \left( b'_0 \realV + b'_1 \realV  \Delta   + b'_2  \Delta \realV  + b'_3  \Delta \right)~,
\label{eq:LH_charged_new}
\eea
and the $b^{(\prime)}_i$ are flavour-blind coefficients.
As explicitly given in~\cite{D'Ambrosio:2002ex,Buras:2010mh}, the $a_i$ and $b^{(\prime)}_i$ 
depend on the $\epsilon^{(\prime)}_i$, on $\tan\beta$, and on the overall normalization
of the Yukawa couplings. Even if  $\epsilon^{(\prime)}_{i}\ll 1$,
the $a_i$ and $b_i$ can reach values of $\cO(1)$ at large $\tan\beta$ and can
be complex, since we allow flavour-blind phases in the model.

\subsection{$\BR(B\to \tau \nu)$ and other observables}

We present here the theoretical expressions of $\BR(B\to \tau \nu)$ 
and a series of other flavour-violating observables,
necessary to set bounds on the parameter space, 
in the $\text{2HDM}_{\overline{\text{MFV}}}$ framework. 

In order to simplify the notations, we absorb terms proportional 
to the top  and bottom Yukawa coupling into the definition of 
$\epsilon_{1,2}^{(\prime)}$. More explicitly, we redefine
$\epsilon_{1,2}^{(\prime)}$ as follows:
\be
\epsilon_{1}^{(\prime)} y_b^2 \to  \epsilon_{1}^{(\prime)}~, \qquad  
\epsilon_{2} ^{(\prime)}y_t^2 \to  \epsilon_{2}^{(\prime)}~. 
\ee
With such a notation, the $b_R \to u_L$
and  $s_R \to u_L$ interactions
with the physical charged Higgs are  
\bea
\cL^{b,s \to u} &=& \frac{m_b \tan\beta}{v} V_{ub}
\frac{1}{1 +(\epsilon_0+\epsilon_1)\tan\beta} \bar u_L  b_R H^+ \no \\
&+& \frac{m_s \tan\beta}{v} V_{us} \frac{1}{1 +\epsilon_0\tan\beta} \bar u_L s_R H^+ 
 {\rm +h.c.} 
\label{eq:Lcc}
\eea
This allows us to derive the following expression for the modification
of $\BR(B\rightarrow \tau \nu)$, relative to the SM, within this framework:
\bea
R_{B\tau\nu} &=& \frac{\BR(B\rightarrow \tau \nu)}{\BR^{\rm SM}(B \rightarrow \tau \nu)}
\no \\
&=& \left[1-\frac{m_B^2}{m_H^2}\frac{\tan^2\beta}{1+(\epsilon_0+\epsilon_1)\tan\beta}
\right]^2~,  \label{eq:Btn} 
\eea
A closely related observable which provide a significant constraint on the parameter space 
is $\BR(K \rightarrow \mu \nu)$. In this case from (\ref{eq:Lcc}) we find
\bea
R_{K\mu\nu} &=&
\frac{\BR(K \rightarrow \mu \nu)}{\BR^{\rm SM}(K \rightarrow \mu \nu)} \no \\
&=& \left[1-\frac{m_K^2}{m_H^2}\frac{\tan^2\beta}{1+\epsilon_0 \tan\beta}\right]^2~.
\eea

Beside semileptonic charged currents, stringent constraints on the  
 $\text{2HDM}_{\overline{\text{MFV}}}$  parameter space 
are provided also by the flavour-changing neutral-current 
(FCNC) transitions $B\rightarrow X_s \gamma$
and $B_s\rightarrow \mu^+\mu^-$. In principle, also the $B_{s}$--$\bar B_{s}$
mixing amplitude could be used to constrain the parameter
space of the model; however, as we will discuss below, it turns out that 
$B_{s}$--$\bar B_{s}$ constraints are automatically satisfied after 
imposing the bounds from $B_s\rightarrow \mu^+\mu^-$.
In order to implement these bounds, we introduce the 
FCNC Hamiltonian 
\be
\cL^{b \to s} = - \frac{G_{\rm F} \alpha_{\rm em}}{
2\sqrt{2}\pi \sin^2 \theta_{\mathrm{W}}}
 V^*_{tb} V_{ts} \sum_{n} C_n {\cal Q}_n~+ ~{\rm h.c.}~,
\ee
where 
\bea
 {\cal Q}_{7} &=& \frac{e}{g^{2}}  m_b \, \bar s
 \sigma_{\mu\nu} \left(1+\gamma_5\right) b
\, F_{\mu\nu}~,  \\
 {\cal Q}^\mu_{S} &=&  
\bar s \left(1+\gamma_5\right) b \; 
\bar \mu \left(1-\gamma_5\right) \mu~,
\eea
and the complete list of effective operators 
can be found in~\cite{Hurth:2008jc}. Following Ref.~\cite{Hurth:2008jc}, 
the experimental constraints on $B\rightarrow X_s \gamma$
and $B_s\rightarrow \mu^+\mu^-$ can effectively be 
encoded into constraints on $C_{7}$ and $C_{S}^\mu$.
More precisely, we can translate the experimental data 
into bounds on $\delta C_{7} = C_{7} (M_W^2) -  C^{\rm SM}_{7} (M_W^2)$
and  $\delta C_{S}^\mu = C_{S}^\mu (M_W^2) -  C_{S}^{\mu \rm SM} (M_W^2)$. 

Working under the hypothesis that the only relevant 
non-standard contributions are those associated to 
the heavy Higgs fields, the dominant contributions to 
$\delta C_{7}$ are the one-loop 
contributions from both charged and neutral Higgs exchange.
Adopting to our notations the 
results of Ref.~\cite{D'Ambrosio:2002ex} we have 
\bea
\delta C_{7}  &=&
\frac{1}{D_{012}}\left[1+(\epsilon'_0+\epsilon'_2)\tan\beta-\frac{\epsilon_2\epsilon'_1\tan^2\beta}{D_{01}}\right]F_7(x^2_{tH})\nonumber\\
&-& \frac{\epsilon_2\tan^3\beta}{D_{012}^2 D_{01}} \frac{x^2_{bH}}{36}
\label{eq:dC7-mod}
\eea
where $x_{qH}=m_q^2/m_H^2$, 
\bea
D_{012}&=&1+(\epsilon_0+\epsilon_1+\epsilon_2)\tan\beta~, \no \\
D_{01}&=&1+(\epsilon_0+\epsilon_1)\tan\beta~,
\eea
and $F_7(x)$ is defined as in~\cite{D'Ambrosio:2002ex}.

The effective coupling of ${\cal Q}^\mu_{S}$ receives contributions
from the FCNC component of (\ref{eq:LH_FCNC}) already at the tree-level:
\be
 \delta C^\mu_S=\frac{m_b m_\mu}{m_H^2} \frac{2\pi \sin^2 \theta_W}{ \alpha_{\rm em} }
 \frac{\epsilon_2 \tan^3\beta}{D_{01} D_{012}}~.
\ee

\section{Phenomenological analysis}

We are now ready to analyse the parameter space of the model,
searching for regions where $\BR(B\rightarrow \tau \nu)$ is enhanced 
over its SM prediction and the other low-energy constraints 
are satisfied. Since the main observables used in CKM fits 
receive tiny corrections from the extended Higgs sector, 
we assume that the standard CKM determination remains valid.

The low-energy phenomenological  constraints used in our analysis are 
\bea 
R_{K\mu\nu}  &\in& (0.98, 1.02)~,   \no \\
\delta C_{7} &\in&  (-0.14,0.06)~, \no \\
\delta C^\mu_S   &\in& (-0.09, 0.09)~.
\label{eq:exp1}
\eea
The first input follows from the analysis of semileptonic $K$ decays in~\cite{Antonelli:2010yf}
(see also~\cite{Colangelo:2010et}),
while the range of $\delta C_{7}$ and $\delta C^\mu_S$ follows from the analysis of  
$B\rightarrow X_s \gamma$ and $B_s\rightarrow \mu^+\mu^-$
performed in~\cite{Hurth:2008jc}. Moreover, since we are interested in 
substantial enhancements of $\BR(B\rightarrow \tau \nu)$,
we impose 
\be
R_{B\tau\nu}  > 1.2~.
\label{eq:exp2}
\ee
On the other hand, given the condition $\epsilon^{(\prime)}_{i} \ll 1$
expected from an approximate $U(1)_{\rm PQ}$ symmetry, 
we will restrict the free parameters of the model to vary in 
the following interval:
\bea
 \tan\beta &\in& (40,60)~, \no \\
m_H[{\rm GeV}] &\in& (150, 1000)~, \no \\
\epsilon^{(\prime)}_i \tan\beta &\in& (-2,2)~.
\label{eq:range}
\eea

\subsection{Analytical considerations}

In principle the model has enough parameters that allow us to 
to satisfy the three conditions in Eqs.~(\ref{eq:exp1}) and, at the same time, get the desired enhancement in $B\to\tau\nu$, 
provided we properly tunes the values of $\epsilon^{(\prime)}_i \times \tan\beta$. 
However, we are not interest in fine-tuned solutions. In particular, 
while it is natural setting to zero some of the $\epsilon^{(\prime)}_i$, 
which are symmetry breaking terms, we consider not natural 
fine-tuned solutions corresponding to 
large values of $\epsilon^{(\prime)}_i \tan\beta$.  
In this perspective, taking into account  the theoretical expressions for the  observables 
presented in the previous section, we find that:
\begin{description}
\item[{\bf \ i.}] Since $\delta C^\mu_S \propto \epsilon_2$, the bound from $B_s\rightarrow \mu^+\mu^-$
can 
easily be satisfied  assuming $\epsilon_2 \approx 0$. This ``natural''  tuning (according 
to the discussion above) allow us to decouple 
charged-Higgs and neutral-Higgs  flavour-changing amplitudes. 
Incidentally, this is why we do not get additional significant
constraints from $B_{s}$--$\bar B_{s}$ mixing.
\item[{\bf \ ii.}] Contrary to $B_s\rightarrow \mu^+\mu^-$,  we cannot get rid of  the $B\to X_s \gamma$ 
bound without some amount of fine tuning. In particular, setting $\epsilon_2 \approx 0$, the charged-Higgs contribution 
to  $B\to X_s \gamma$ vanishes completely only under the fine-tuned condition
\be
(\epsilon'_0+\epsilon'_2)\tan\beta = -1~.
\label{eq:finetune}
\ee 
\end{description}

Before analysing how far from the fine-tuned condition in Eq.~(\ref{eq:finetune}) we can move,
it is worth discussing the correlation between $B\to\tau\nu$ and $K\to\mu\nu$
ignoring all other constraints. 

In the case of $K\to\mu\nu$, the Higgs-mediated amplitude is always
much smaller that the SM one. Imposing $R_{K\mu\nu}\in [0.98,1.02]$ implies
\be
\left|\epsilon _0 \tan \beta
   +1\right|> 0.9 \times  r 
   \label{eq:eqkn1}
   \ee
where $r = m_{B}^2 \tan^2\beta  /m_H^2 $.  
For the chosen range of $\tan \beta$ and $m_H $ we have 
$0.04  <  r <  4.3$. For small values of $r$ the above condition is 
very natural: we only exlude a narrow region  around the (unnatural)
point $\epsilon_0\tan\beta = -1$.  On the other hand, 
for growing values of $r$ we are pushed toward a fine tuned
configuration. We thus conclude that the $K\to\mu\nu$ bound
points toward small values of  $r$.

Two solution are possible to generate an enhancement of 
$\BR(B\rightarrow \tau \nu)$:
a destructive interference of SM and char\-ged-Higgs amplitudes, 
if the latter is more than twice the SM one in size;
a constructive interference of SM and char\-ged-Higgs amplitudes, 
independently from the size of the charged-Higgs amplitude.
Requiring $R_{B\tau\nu}  > 1.2$ implies 
\be
-(1-\epsilon_0\tan\beta)<\epsilon_1\tan\beta<-(1-\epsilon_0\tan\beta) + 0.5 \times r
 \label{eq:eqbn1}
\ee
for the case of destructive interference, and 
\be
(1-\epsilon_0\tan\beta) < - \epsilon_1\tan\beta < (1-\epsilon_0\tan\beta) + 10 \times r  
 \label{eq:eqbn2}
\ee
for the case of constructive  interference. It is clear from the above equations 
that the constructive case allow a larger region of the parameter space. This is particularly 
true for small values of $r$, as suggested by the $K\to\mu\nu$ bound. As we will
discuss in the following, this conclusion remains true and is even reinforced 
once we take into account also the $B\to X_s \gamma$ bound.  
Finally, a destructive interference 
of scalar and SM amplitudes in $b \rightarrow  c \tau\nu$,
able to increase $\BR(B\rightarrow \tau \nu)$, is strongly 
 disfavored by $B\to D\tau \nu$ data~\cite{Kamenik:2008tj}
 and the lower bounds on $m_H$ from direct searches at the LHC 
 (see discussion below).

We finally comment on previous analyses about the possiblity to enhance $\BR(B\rightarrow \tau \nu)$ in 2HDMs.
A general analysis in the context of the Higgs sector of the minimal supersymmetric extension of the SM (MSSM) has been presented in 
Ref.~\cite{Bhattacherjee:2010ju}. In that framework the  $\epsilon_i$ are not free parameters. As a result, their analysis is less general than the one
presented here, at least as Higgs-mediated amplitudes are concerned. In particular, the constructive interference solution,
occurring for $1+(\epsilon_0+\epsilon_1)\tan\beta<0$ has not been considered in Ref.~\cite{Bhattacherjee:2010ju}. The importance of 
the latter has been pointed out first in Ref.~\cite{Akeroyd:2003zr}. However, in the latter work the correlation with the other observables 
we are considering has not been analyzed.

\begin{figure}[t]
\hspace{0.0 true cm}
\includegraphics[width=0.48\textwidth]{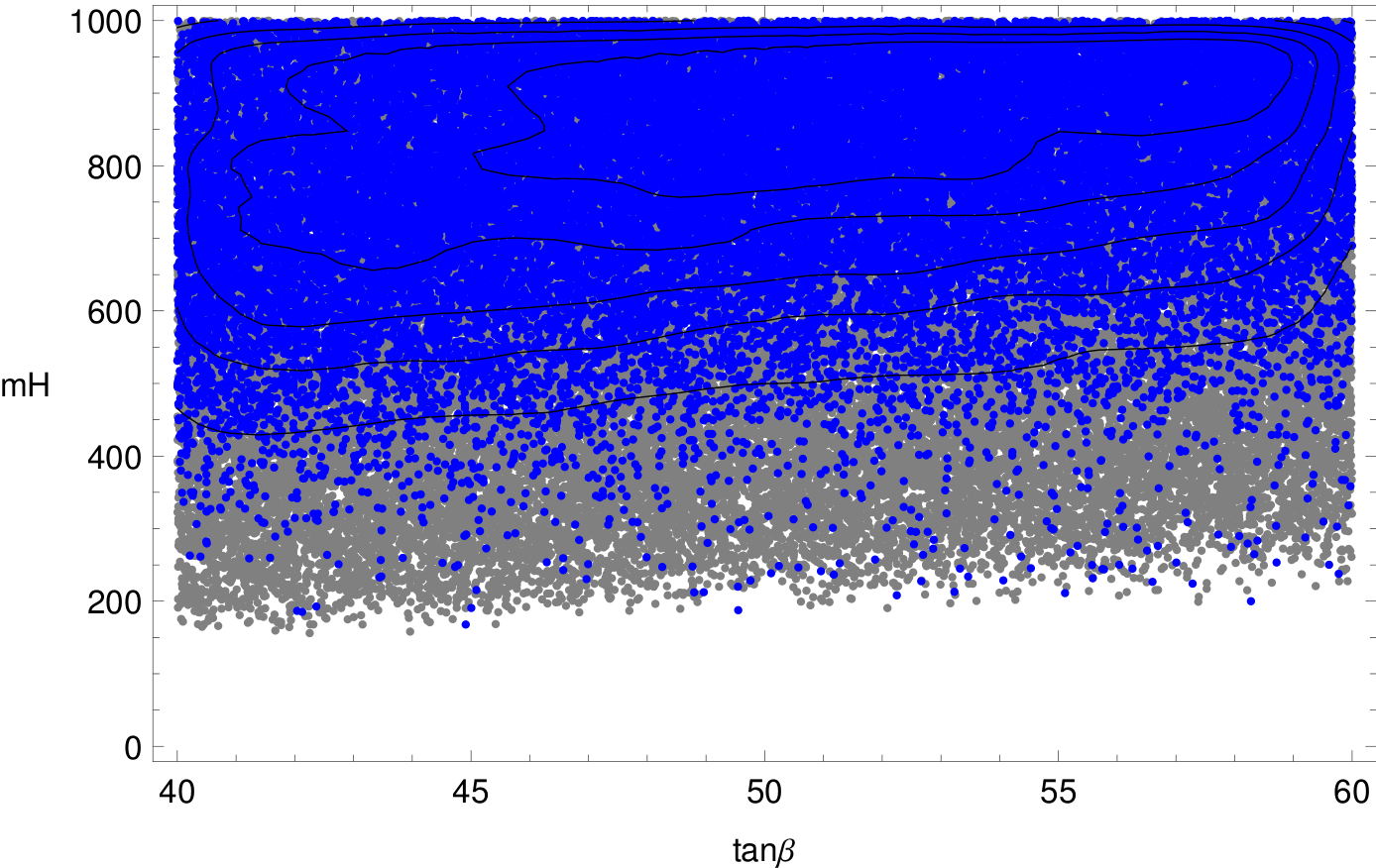}
\caption{\label{fig:mhtb}
Allowed regions in the $\tan\beta$--$m_H$ plane. The grey points 
correspond to regions of the parameter space that can be reached only in the fine-tuned 
configuration where $\epsilon_2=0$ and $\epsilon'_{0,2}$  are fixed to 
satisfy the condition (\ref{eq:finetune}).  The black contours mark equally-populated areas 
resulting from the the global sampling (without fine-tuned conditions): the three most inner contours
include 50\% of the points.}
\end{figure}

\subsection{Numerical analysis}

In order to analyze all the constraints at the same time, trying to avoid fine-tuned configurations,
we have randomly generated values for the relevant $\epsilon_i^{(\prime)} \tan\beta$ using (uncorrelated) Gaussian distributions centered in zero 
--corresponding to the limit of exact $U(1)_{\rm PQ}$  symmetry-- and with $\sigma=0.5$. The values of $m_H$ and $\tan\beta$
are extracted with uniform distributions in the ranges specified in Eq.~(\ref{eq:range}). 

\begin{figure}[t]
\hspace{0.0 true cm}
\includegraphics[width=0.48\textwidth]{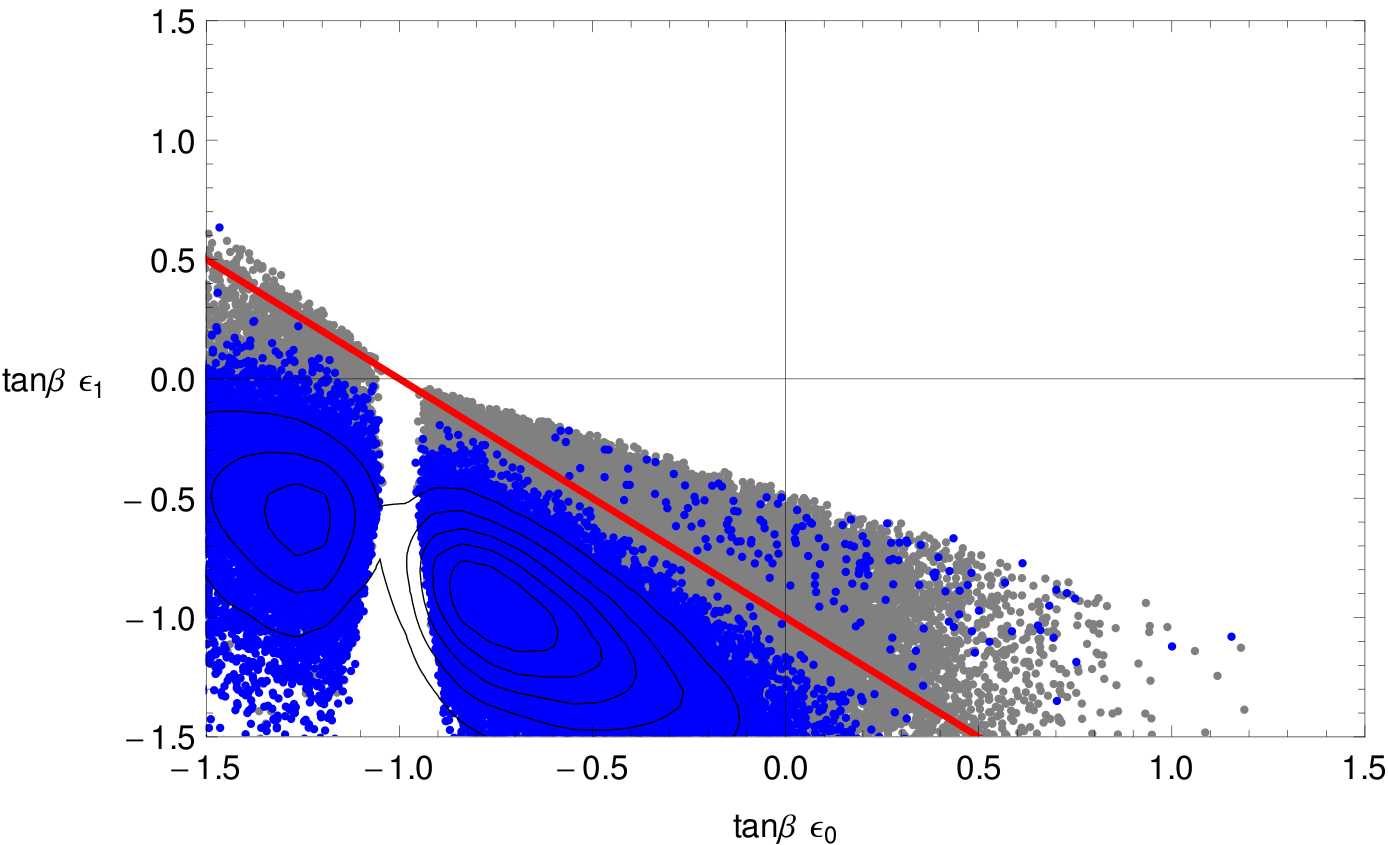}
\caption{\label{fig:e0e1} 
Allowed regions in the $\epsilon_0$--$\epsilon_1$ plane. 
Notations as in Fig.~\ref{fig:mhtb}. }
\end{figure}

\begin{figure}[t]
\hspace{0.0 true cm}
\includegraphics[width=0.48\textwidth]{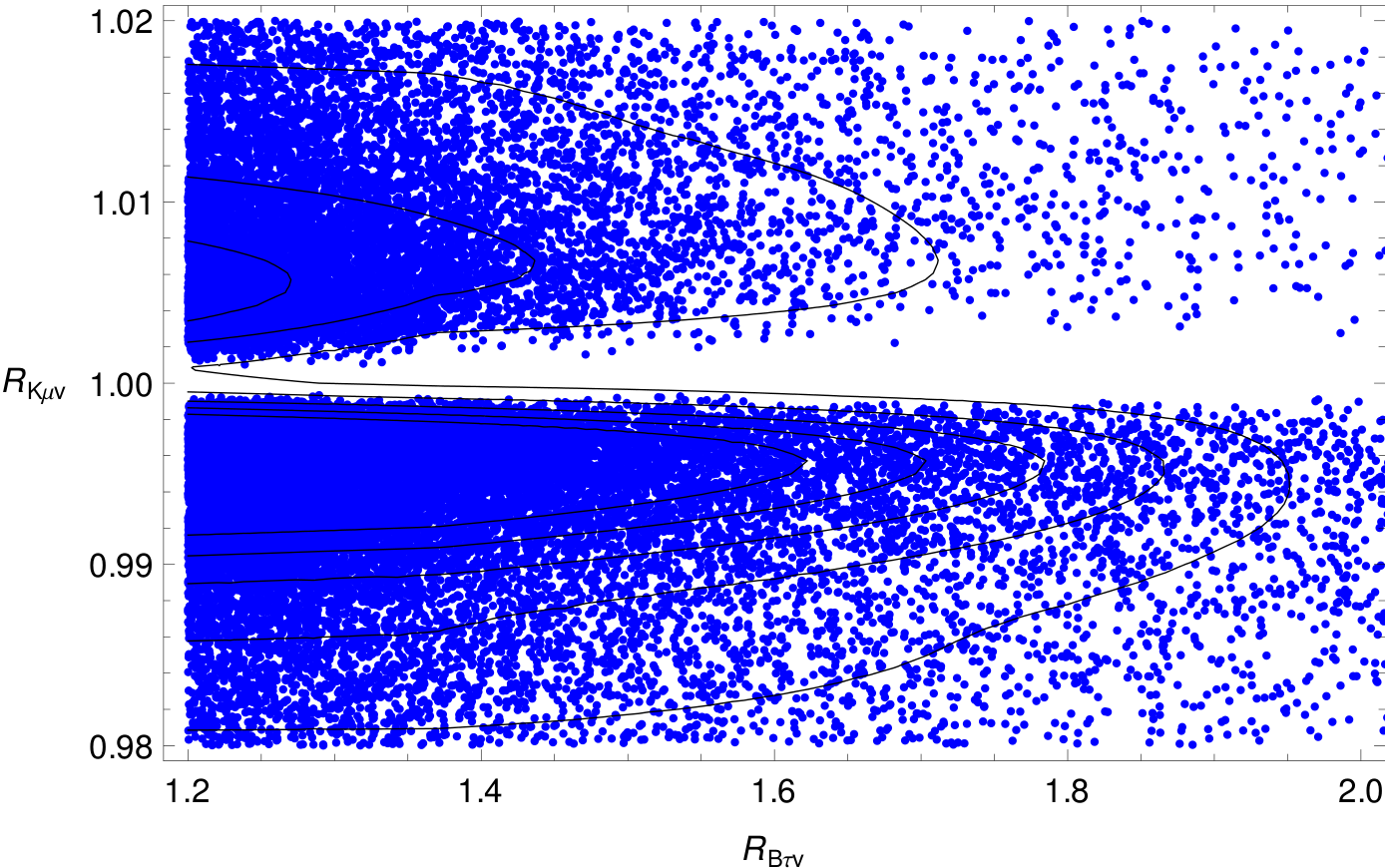}
\caption{\label{fig:BvsK} 
Allowed regions in the $R_{B\tau\nu}$--$R_{K\mu\nu}$ plane. }
\end{figure}

\begin{figure}[t]
\hspace{0.0 true cm}
\includegraphics[width=0.48\textwidth]{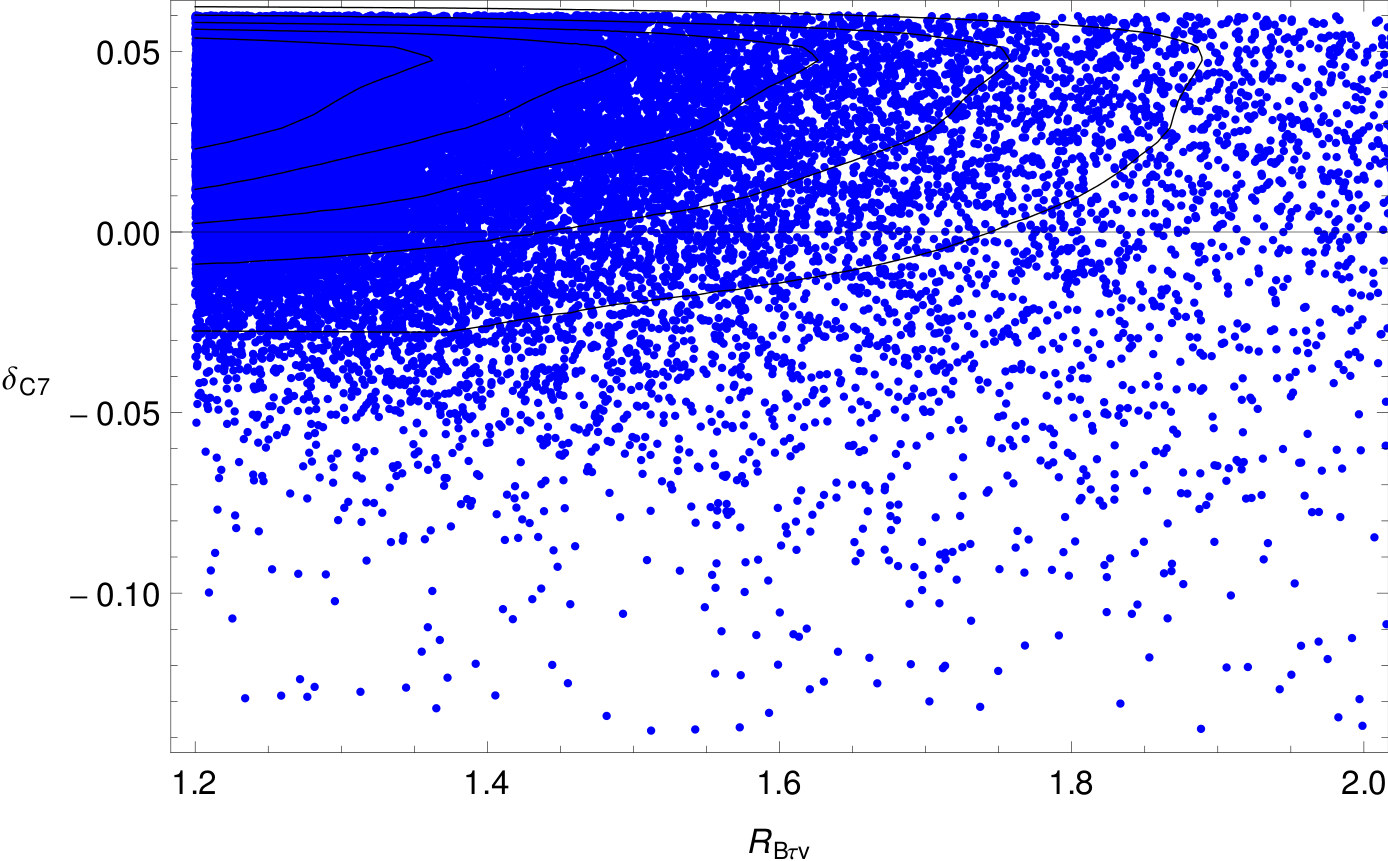}
\caption{\label{fig:dC7} 
Allowed regions in the $R_{B\tau\nu}$--$\delta C_7$ plane.  }
\end{figure}

The results of this numerical analysis are shown in Fig.~\ref{fig:mhtb}--\ref{fig:dC7}.  In Fig.~\ref{fig:mhtb} and 
Fig.~\ref{fig:e0e1} we show the points satisfying all constraints in Eqs.~(\ref{eq:exp1})--(\ref{eq:exp2}).
To better quantify the role of the $B\to X_s \gamma$ and $B_s\rightarrow \mu^+\mu^-$ bounds, 
we have plotted with a different color (grey point) the region of parameter space that can be reached only in the fine-tuned 
configuration where $\epsilon_2=0$ and $\epsilon'_0$ and $\epsilon'_2$  are fixed to 
satisfy the condition (\ref{eq:finetune}). As can be seen from Fig.~\ref{fig:mhtb}, 
in general there is no significant constraint on $m_H$ and $\tan\beta$; however, low values 
of  $m_H$ can be obtained only in the in the fine-tuned configuration. 

At this point it is worth to comment on the bounds in the $m_H$--$\tan\beta$ plane 
by direct searches for heavy Higgs bosons at the LHC~\cite{Chatrchyan:2012vp}. A
direct implementation of these constraints in our frameworks is not possible, given the
former are obtained in the limit limit $\epsilon_i^{(\prime)}=0$. Still, it is worth to note that in this limit 
direct searches set the approximate bound $m_H \gsim 420~{\rm GeV}+ 6 \times (\tan\beta - 40)$~\cite{Chatrchyan:2012vp},
which does not represent a problem for most of the points in Fig.~\ref{fig:mhtb}.
Only the fine-tuned (gray) points are potentially affected by this constraint, which thus provide a 
further argument against the tuned configuration with low $m_H$.

In Fig.~\ref{fig:e0e1} we show the  points satisfying all constraints in the $\epsilon_0$--$\epsilon_1$ plane.
We also show the line $1+(\epsilon_0+\epsilon_1)\tan\beta = 0$, separating the region
of destructive interference (above the line) and constructive interference (below the line) in $B\to\tau\nu$.
As can be seen, the region of constructive interference is reached  essentially only in the fine-tuned 
configuration where  $\epsilon_2$ and $\epsilon'_{0,2}$ are fixed to eliminate any non-standard 
contribution to $B\to X_s \gamma$ and $B_s\rightarrow \mu^+\mu^-$. Indeed in this region we need large
values of $m_H$, that would get in conflict with $B\to X_s \gamma$ and $B_s\rightarrow \mu^+\mu^-$ 
for generic values of $\epsilon_2$ and $\epsilon'_{0,2}$. On the other hand, the region of constructive interference
is densely populated even in absence of a fine-tuning on $\epsilon_2$ and $\epsilon'_{0,2}$.
As anticipated in the analytical discussion, the absence of points for  $\epsilon_0\tan\beta $ close to -1
is a consequence of the $K\to\mu\nu$ bound. Last but not least, we stress the absence of points near 
the  $U(1)_{\rm PQ}$  symmetric point  $\epsilon_0 = \epsilon_1 = 0$.
This is a simple consequence of combining the  $K\to\mu\nu$ and $B\to\mu\nu$ constraints 
in Eqs.~(\ref{eq:eqkn1})--(\ref{eq:eqbn2}).

In Fig.~\ref{fig:BvsK} and~\ref{fig:dC7} we show the correlation of $R_{B\tau \nu}$
with the two most significant constraints, namely  $\BR(K \rightarrow \mu \nu)$ and 
$\delta C_7$ (or $B \to X_s\gamma$) (in these two figures we do not plot with different colors the fine-tuned points).
 As can be seen from Fig.~\ref{fig:dC7}, 
 the interference between SM and charged-Higgs amplitudes 
is necessarily destructive in  $B\to X_s \gamma$ (we recall that $C_7^{\rm SM} <0$).
On the other hand, both positive and negative interferences in  $K \rightarrow \mu \nu$
are possible, depending on the sign of $1+\epsilon_0\tan\beta$.
As illustrated in  Fig.~\ref{fig:BvsK}, if the maximal deviation 
from the SM in $\BR(K \rightarrow \mu \nu)$ could be reduced to $1\%$,
the parameter space leading to an 
enhancement of $\BR(B \rightarrow \tau \nu)$ would be strongly reduced.
This also implies that if the precision on $\BR(K \rightarrow \mu \nu)$ will improve,
there are realistic chances to see a deviation from the SM in this mode within this
framework.  On the contrary, we have checked that for $R_{B\tau \nu} < 2$
 the deviations from the SM predictions  in $\BR(B \to D\tau \nu)$ 
 do not exceed the $20\%$ level, well within the present 
 theoretical and experimental uncertainties~\cite{Kamenik:2008tj}.
 
 As a final check of the stability of our findings, we have performed
 scan of the parameter space allowing arbitrary complex phases for
 the $\epsilon_i^{(\prime)}$. As expected, no significant deviations 
 in Fig.~ \ref{fig:mhtb}, \ref{fig:BvsK} , and \ref{fig:dC7} has been 
 observed. Fig.~\ref{fig:e0e1} is unaffected provided we interpret it 
 as the  Re($\epsilon_0$)--Re($\epsilon_1$) plane.

\section{Discussion and Conclusions}

Our analysis shows that is possible to 
accommodate sizable enhancements of $\BR(B \rightarrow \tau \nu)$
in the $\text{2HDM}_{\overline{\text{MFV}}}$ framework, 
despite the tight constraints of other low-energy 
observables. This is clearly illustrated by the plots discussed
in the previous section.  
However, it must be stressed that this enhancement occurs
under a few  specific circumstances:
\begin{description}
\item[{\bf \ i.}] At least some of the  $\epsilon_i\tan\beta$
must of order one, i.e.~sizable deviations from the exact 2HDM-II limit,
or from the limit of unbroken $U(1)_{\rm PQ}$ symmetry in the Yukawa sector, 
are necessary. As shown in Fig.~\ref{fig:e0e1}, almost no solution survive 
for $|\epsilon_i\tan\beta| < 0.5$. This conclusion holds independenlty of the 
simplifying assumptions on the $\epsilon_i$ adopted in the 
present analysis. 
\item[{\bf \ ii.}] 
If we assume $\epsilon_i \ll 1$, as realised in several explicit models
where the $U(1)_{\rm PQ}$ symmetry in the Yukawa sector is 
broken only by radiative corrections, the need for large $\epsilon_i\tan\beta$ 
necessarily imply large $\tan\beta$ values. 
\item[{\bf iii.}] In addition of being sizable, the 
$U(1)_{\rm PQ}$ breaking terms $\epsilon_{0}$ and $\epsilon_{1}$
should conspire to suppress the 
combination  $1+(\epsilon_0+\epsilon_1)\tan\beta$ 
appearing in the denominator of the $B \rightarrow \tau \nu$
amplitude. The more $m_H$ is large, the more fine tuning 
on $1+(\epsilon_0+\epsilon_1)\tan\beta$ is needed in order 
to keep the charged-Higgs amplitude at the level of the SM one.
\item[{\bf iv.}] The  most likely possibility to enhance $B \rightarrow \tau \nu$,
especially if  $m_H$ is above 200 GeV, occurs in the case of constructive
interference between SM and charged Higgs amplitudes in $B \rightarrow \tau \nu$.
This requires value of the $\epsilon_i$ which cannot be obtained in 
simplified MSSM scenarios, such as the one considered in Ref.~\cite{Bhattacherjee:2010ju},
but can be obtained in less standard supersymmetric frameworks, such as the 
"up-lifted"  scenario considered in Ref.~\cite{Dobrescu:2008sz}.
\end{description}
If the above conditions are satisfied, a large enhancement 
of  $\BR(B \rightarrow \tau \nu)$ is compatible with the 
existing constraints. In absence of fine tuning, this implies 
non-negligible and potentially visible deviations from the SM in  $B\to X_s \gamma$ and 
$K \rightarrow \mu \nu$. The most interesting effects are expected 
in $\BR(K \rightarrow \mu \nu)$, as illustrated in 
Fig.~\ref{fig:BvsK}. To this purpose, we stress that 
$\BR(K \rightarrow \mu \nu)$ is presently measured with 
a $0.27\%$ relative error~\cite{Ambrosino:2005fw}. 
If future lattice determinations 
of the kaon form factors could allow us to reduce the 
theoretical error on $\BR(K \rightarrow \mu \nu)$
at the same level, the $\BR(B \rightarrow \tau \nu)$--$\BR(K \rightarrow \mu \nu)$
correlation would provide a useful tool to 
test this framework.

\section*{Acknowledgments}
We thank Stefania Gori for useful comments.
This work was supported by the EU ERC Advanced Grant FLAVOUR (267104),
and by MIUR under contract 2008\-XM9HLM.
G.I. acknowledges the support of the Technische Universit\"at M\"unchen 
-- Institute for Advanced Study, funded by the German Excellence Initiative.

\end{document}